\documentclass[10pt]{article}
\usepackage{psfig,epsf,conf-X}

\begin{document} 
\small
\heading{%
%
The X-ray/submillimetre Link
}
\par\medskip\noindent
\author{%
Omar Almaini
}
\address{%
Institute for Astronomy, Royal Observatory, University Of Edinburgh, 
Blackford Hill, Edinburgh EH9 3HJ, Scotland
}

\begin{abstract}
It is widely believed that most of the cosmic X-ray background (XRB)
is produced by a vast, hitherto undetected population of obscured
AGN. Deep X-ray surveys with Chandra and XMM will soon test this
hypothesis. Similarly, recent sub-mm surveys with SCUBA have revealed
an analogous population of exceptionally luminous, dust-enshrouded
{\em star-forming} galaxies at high redshift.  There is now growing
evidence for an intimate link between these obscured populations.
There are currently large uncertainties in the models, but several
independent arguments lead to the conclusion that a significant
fraction of the SCUBA sources ($10-30\% $) will contain
quasars. Recent observational studies of SCUBA survey sources appear
to confirm these predictions, although the relative roles of AGN and
star-forming activity in heating the dust are unclear.  Forthcoming
surveys combining X-ray and sub-mm observations will provide a very
powerful tool for disentangling these processes.

\end{abstract}
\section{Obscured AGN and the X-ray background}

The origin of the cosmic X-ray background (XRB) has been a puzzle for
over 35 years, but now there is strong evidence that this will be
explained by a large population of absorbed AGN. If these are to
explain the bump in the XRB spectrum at 30keV, however, the total
energy output of this population must exceed that of broad-line AGN by
at least a factor of $\sim 5$, with wide ranging implications
\cite{fabian98}.  This hidden population could also explain the
apparent discrepancy between the black hole densities predicted by
ordinary QSOs and recent observations of local galaxy bulges
\cite{fi99}, \cite{magorrian98}.

Deep X-ray surveys with Chandra and XMM will soon test this obscured
AGN hypothesis, although existing  surveys with ROSAT, ASCA and
BeppoSAX have already revealed what could be the `tip of the iceberg'
of this population.  Several unambiguous cases of obscured QSOs have
been detected (e.g. \cite{oa95}, \cite{ohta96}, \cite{iwasawa97},
\cite{bjb98}, \cite{barcons98}) while at the faintest X-ray fluxes
there is growing evidence for a large population of X-ray luminous
emission-line galaxies, many of which show clear evidence for AGN
activity  \cite{bjb95},\cite{carballo95},\cite{oa97},\cite{hardy98} .

Arguably the most convincing evidence for the obscured AGN model came
from the ultra-deep survey of Hasinger, Schmidt et al
(\cite{hasinger98}, \cite{schmidt98}). Using the Keck telescope to
identify sources from the deepest X-ray observation ever taken they
found that most of these X-ray galaxies could be classified as
AGN. Exciting new work has also been undertaken at harder energies
with ASCA and Beppo-SAX (\cite{ig97}, \cite{fiore99},
\cite{akiyama98}) resolving $\sim 30\% $ of the $2-10$\,keV XRB.
Increasing numbers of these sources have been identified with absorbed
AGN.

\section{Star-forming Galaxies  and Submillimetre Surveys}

Deep sub-mm observations offer the potential to revolutionise our
understanding of the high redshift Universe.  Beyond 100$\mu$m, both
starburst galaxies and AGN show a very steep decline in their
continuum emission, which leads to a large negative K-correction as
objects are observed with increasing redshift. This effectively
overcomes the `inverse square law' to pick out the most luminous
objects in the Universe to very high redshift \cite{blain93}.  Since
the commissioning of the SCUBA array at the James Clerk Maxwell
Telescope a number of groups have announced the results from deep
sub-mm surveys, all of which find a high surface density of sources at
$850 \mu$m ( \cite{smail97}, \cite{hughes98}, \cite{barger98},
\cite{eales99}, \cite{blain99}). The implication is the existence of a
large population of hitherto undetected dust enshrouded galaxies.  In
particular, the implied star-formation rate at high redshift ($z>2$)
is significantly higher than that deduced from uncorrected optical-UV
observations, roughly a factor of two higher than even the
dust-corrected version of the optically derived star-formation history
\cite{blain99}. The recent detections of the far-infrared/sub-mm
background by the DIRBE and FIRAS experiments provide further
constraints, representing the integrated far-infrared emission over
the entire history of the Universe (\cite{puget96}, \cite{fixsen98},
\cite{hauser98}).  Since most of this background has now been resolved
into discrete sources by SCUBA, the implication is that most high
redshift star-forming activity occurred in rare, exceptionally
luminous systems.

In other words, at high redshift ULIRG-like starburst galaxies
dominate the cosmic energy budget, in stark contrast to the situation
today where (to quote Andy Lawrence) `ULIRGs are little more than a
spectacular sideshow' \cite{al99}.

\section{The X-ray/sub-mm link}

Considerable excitement has been generated recently by the possibility
that many of these SCUBA sources could be AGN.  Whether these AGN are
actually heating the dust is another matter (see Section 5) but there
are now several independent lines of argument which suggest that AGN
are present in a significant fraction of these SCUBA sources.  At the
very least, the implication is that much of the star formation in the
high redshift Universe occurred in galaxies containing active
quasars. First we present the arguments predicting an AGN
contribution, followed by recent observational evidence.

\subsection{Arguments for AGN in deep sub-mm surveys}

\begin{itemize}

\item{{\bf The analogy with ULIRGs:} In many ways a significant AGN
fraction would not be a surprise. The SCUBA sources are exceptionally
luminous systems, essentially the high redshift equivalents to local
ULIRGs. At the luminosities of the SCUBA sources ($\sim
10^{12}L_{\odot}$) we note that a least $30\% $ of local ULIRGs show
clear evidence for an AGN \cite{sm96}.}

\item{{\bf Predictions based on AGN luminosity functions:} One can
estimate the AGN contribution by transforming the X-ray luminosity
function to the sub-mm waveband using a template AGN SED. We estimate
that $10-20\% $ of the sources in recent SCUBA surveys could contain
AGN, perhaps higher if one allows for Compton-thick objects
\cite{oa99}.  Major sources of uncertainty are in the extrapolation of
local AGN SEDs to high redshift, the dust temperature and in assuming
the same underlying luminosity function for obscured AGN.  Note that
an independent but very similar analysis by Manners et al (in
preparation) predicts a lower fraction ($5-10 \% $).}

\item{{\bf Re-radiating the absorbed energy:} Another approach, also
based on obscured AGN models for the XRB, does not rely on uncertain
SEDs but instead on thermally re-radiating the absorbed energy
directly in the far-infrared \cite{gunn99}. This method also predicts
a significant AGN fraction among the SCUBA sources ($5-30\% $)
although the exact prediction is strongly dependent on the assumed
dust temperature.}

\item{{\bf SCUBA observations of high-z quasars:} Observations of the
most luminous, very high redshift quasars ($z>3$) suggest that many
are exceptionally luminous in the far-infrared/sub-mm, with sub-mm
luminosities comparable to Arp 220 \cite{mcmahon99}.  Whether this
emission is due to dust heated by the quasar or associated with
starburst activity is unclear, but the lack of any correlation between
the quasar power and the sub-mm luminosity would favour a starburst
origin for the far-infrared emission. If one assumes that all high
redshift quasars have similar sub-mm luminosities this would lead to a
very large AGN fraction in the deep SCUBA surveys ($\sim 50 \% $). In
reality, however, some weak correlation between quasar power and the
associated starburst is likely to exist, which would reduce this
fraction significantly.  Further sub-mm observations of high-z quasars
are required to investigate these correlations.}

\end{itemize}

\subsection{Direct evidence for AGN in sub-mm surveys}

\begin{itemize}

\item{{\bf The SEDs of detected SCUBA sources}: A recent analysis of
the multi-wavelength spectral energy distributions (SEDs) of SCUBA
sources has suggested that $\sim 1/3$ are likely to be AGN
\cite{cooray99}. }

\item{{\bf Spectroscopic identification:} Although the identification
of many SCUBA deep survey sources remains elusive (perhaps indicating
their exceptionally high redshift) there are growing indications that
a significant fraction harbour AGN.  The first clear-cut
identification turned out to be an obscured QSO \cite{ivison98} and
since then various surveys have been able to place limits on the AGN
fraction.  From a study of $14$ sub-mm sources, Barger and
collaborators have placed a lower limit of $\sim 20 \% $ on the
fraction showing evidence for AGN activity \cite{barger99}. Of seven
sub-mm sources studied in detail by Ivison et al, at least $3$ show
evidence for an AGN \cite{ivison99}. These estimates are in good
agreement with the predictions of the various models outlined above.}

\end{itemize}

\section{Probing the X-ray/sub-mm link with Chandra and XMM}

Forthcoming deep X-ray surveys with Chandra and XMM will push
significantly fainter than ever before.  In the soft X-ray band, we
expect to reach at least an order of magnitude fainter than the
deepest ROSAT surveys, while in the hard X-ray band the improvement is
even more dramatic (Figure 1). The puzzle of the XRB should therefore
soon be solved, and we expect to detect large numbers of obscured AGN
and study their properties in detail. In addition, we will be able to
detect typical quasars to very high redshift ($z\sim8$) and hence
assess the importance of quasar activity during those early epochs.

These deep X-ray observations are potentially ideal for identifying
AGN in sub-mm surveys. The source densities expected are very similar
($\sim 1000$ deg$^{-2}$) and hence with the resolution of Chandra in
particular it will be possible to pick out the AGN directly from their
X-ray flux, with very little confusion.

The $8$mJy survey of the UK SCUBA consortium is ideal for such a study
(Figure 2).  This is the only wide area SCUBA survey being conducted,
covering $500$ square arcminutes in $2$ contiguous regions. One of
these regions is in the Lockman Hole (which will be covered by Chandra
in PV time). The other is concentrated on the N2 region of the ELAIS
survey, where we have deep Chandra and XMM observations planned (PI:
Almaini).  We will be able to detect the hard X-ray emission from the
hidden AGN {\em and} obtain a measurement of the absorbing column.  If
the equivalent width is high enough, in some cases XMM may even allow
a redshift determination from an Iron line detection

\begin{figure}
\centerline{\vbox{
\psfig{figure=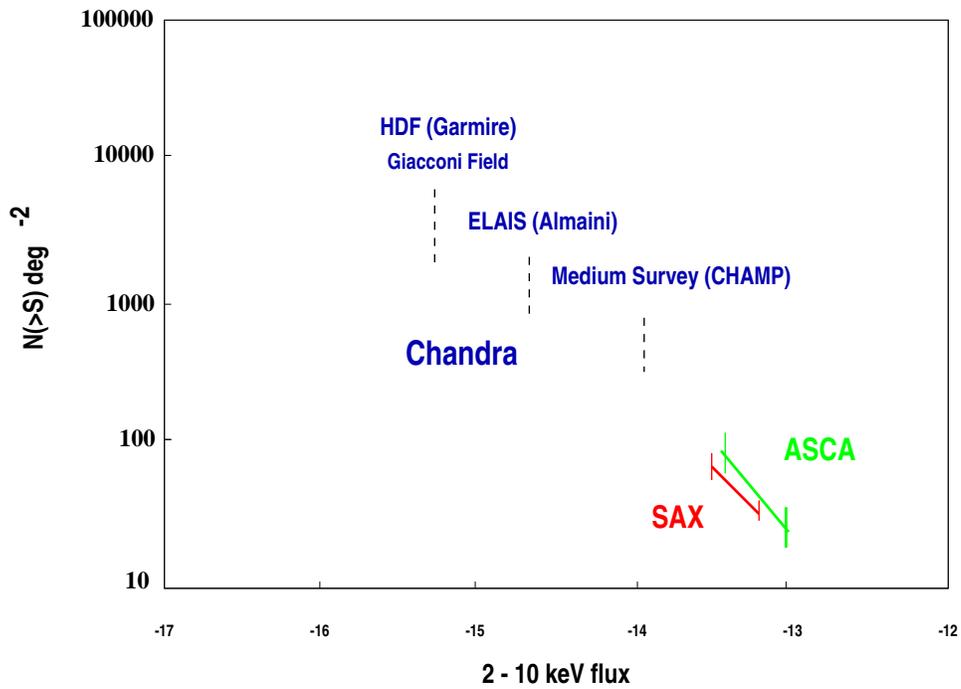,height=9.cm}
}}
\caption[]{Predicted hard X-ray source counts for forthcoming Chandra
surveys, compared with the deepest obtained so far (with ASCA and
BeppoSAX).  
}
\end{figure}

\begin{figure}
\centerline{\vbox{
\psfig{figure=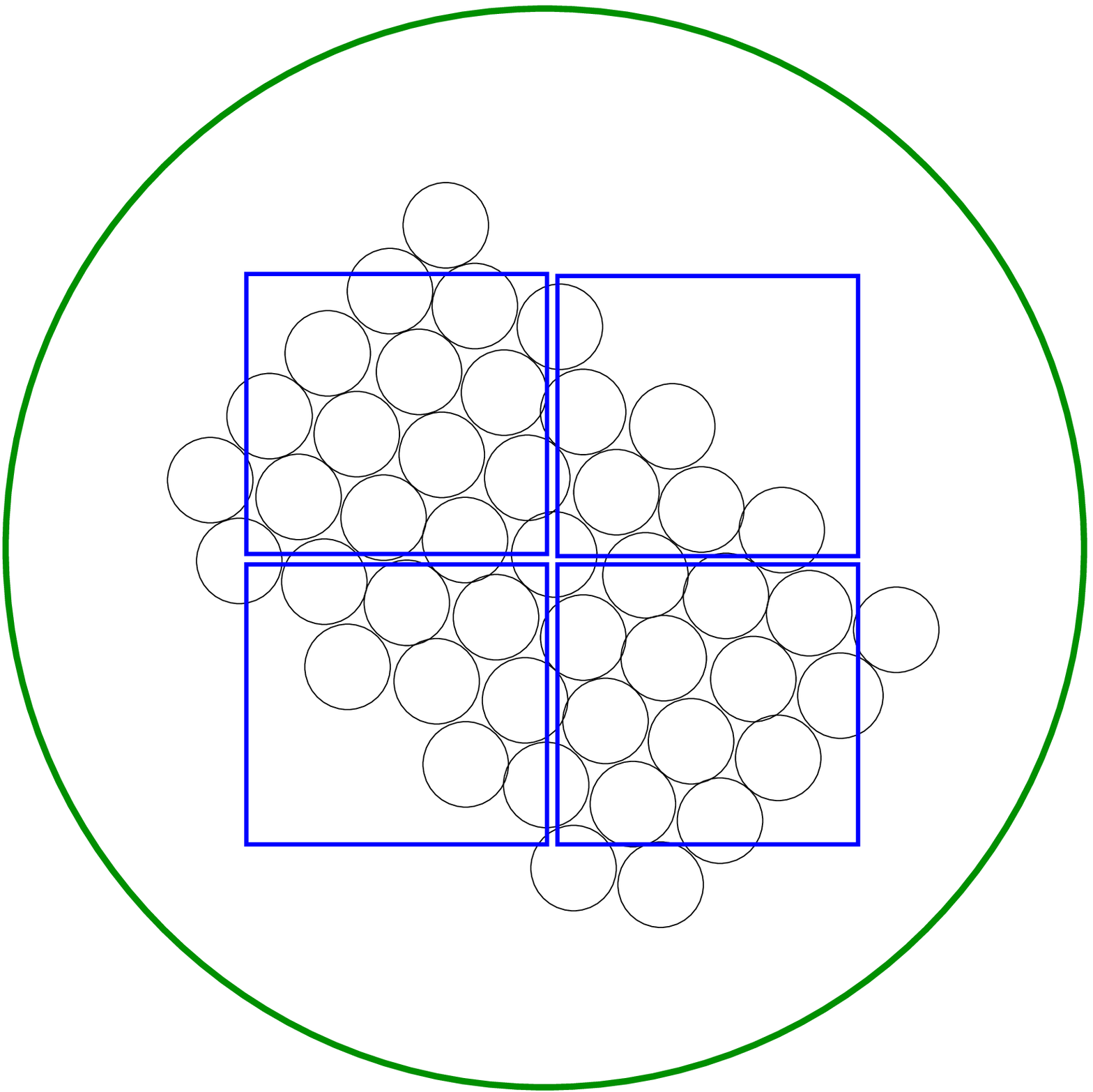,height=11.cm}
}}
\caption[]{Showing the N2 region from the UK SCUBA survey
(small circles) in which we expect to detect $\sim 50$ submillimetre
sources.  This field is being observed for $75$ks by Chandra (squares)
and $150$ks by XMM (large circle) }
\end{figure}

\section{Conclusions and Implications}

Several independent arguments now point to the conclusion that a
significant fraction ($10-30 \% $) of the luminous sub-mm sources
detected by SCUBA will contain AGN. This points to a very important
link between X-ray astronomy and the newly emerging sub-mm field, both
of which provide probes of the obscured, high redshift Universe.

If a significant AGN fraction is confirmed with forthcoming
Chandra/XMM surveys, considerable uncertainties will still remain. Is
the dust heated by the AGN or by stellar processes?  If the AGN is
responsible, and their contribution is large, the recent conclusions
about star-formation at high redshift may require significant
revision.  On the other hand the dust may be largely heated by stellar
activity (see \cite{mrr99}) but with the interesting implication that
much of the star formation at high redshift occurred in galaxies
containing active quasars.  It has recently been postulated that
perhaps all quasars could go through an obscured phase during the
growth of the black hole, a process which could be intimately linked
with the formation of the galaxy bulge itself (see \cite{fabian99}).
Future surveys combining X-ray and sub-mm observations will provide a
powerful tool for disentangling these processes.

\vfill
\end{document}